\documentclass[11pt]{iopart}
\usepackage[latin1]{inputenc}
\usepackage[spanish,english]{babel}
\usepackage[dvips]{graphicx}
\usepackage{color}
\usepackage{amscd}
\usepackage{enumerate}
\usepackage{dsfont}
\usepackage{amssymb}
\usepackage{amsthm}
\usepackage{syntonly}
\usepackage{undertilde}
\usepackage{hyperref}
\usepackage{tensor}
\usepackage{accents}
\usepackage{cite}
\begin{document}
 
\title[Lorentz-covariant Hamiltonian analysis of BF gravity with the Immirzi parameter]{Lorentz-covariant Hamiltonian analysis of BF gravity with the Immirzi parameter}

\author{Mariano Celada and Merced Montesinos}
\address{Departamento de Física, Cinvestav, Av. Instituto Politécnico Nacional 2508, San Pedro Zacatenco, 07360, Gustavo A. Madero, Ciudad de México, México.
}
\ead{mcelada@fis.cinvestav.mx, merced@fis.cinvestav.mx}

\selectlanguage{english}

\begin{abstract}

 We perform the Lorentz-covariant Hamiltonian analysis of two Lagrangian action principles that describe general relativity as a constrained BF theory and that include the Immirzi parameter. The relation between these two Lagrangian actions has been already studied through a map among the fields involved. The main difference between these is the way the Immirzi parameter is included, since in one of them the Immirzi parameter is included explicitly in the BF terms, whereas in the other (the CMPR action) it is in the constraint on the B fields. In this work we continue the analysis of their relationship but at the Hamiltonian level. Particularly, we are interested in seeing how the above difference appears in the constraint structure of both action principles. We find that they both possess the same number of first-class and second-class constraints and satisfy a very similar (off-shell) Poisson-bracket algebra on account of the type of canonical variables employed. The two algebras can be transformed into each other by making a suitable change of variables.

\end{abstract}

%\maketitle

\section{Introduction}

General relativity that was conceived by Einstein as a theory for describing the gravitational field employs a metric tensor as its fundamental variable. It has been also formulated in terms of alternative variables as time has passed. The Einstein-Hilbert action, based on a metric tensor, was the first action principle from which Einstein's equations come from. Despite being a diffeomorphism-invariant action principle, it does not explicitly show all the relevant symmetries behind gravity. Only after the tetrad field was introduced, general relativity became a theory explicitly endowed with two fundamental symmetries: diffeomorphism and local Lorentz invariances, being the latter relevant for the coupling of fermions to gravity. The Hilbert-Palatini action fulfils these requirements. In this formulation, the tetrad field and the Lorentz connection 1-form are the independent variables from which pure general relativity comes from.

Once the Lagrangian action for the gravitational field was achieved, the physicists' efforts turned toward the Hamiltonian form of this action principle as a previous stage required for implementing a canonical quantization programm of gravity. The ADM formulation \cite{Witten} set up the Hamiltonian analysis of the Einstein-Hilbert action showing that general relativity was a constrained theory as expected, but its full quantization was never achieved successfully because of technical and conceptual problems. However, the canonical programm was boosted when Ashtekar \cite{Ash} performed a complex canonical transformation from the ADM variables (in the tetrad formalism) to the so-called Ashtekar variables of complex general relativity, simplifying in this way the constraints of the theory and making possible the use of Yang-Mills techniques for building the Hilbert space of the theory. Nevertheless, it was very difficult to handle the reality conditions required to get real general relativity at the quantum level.

On the other hand, in the 1970s Pleba\'{n}ski \cite{Pleb} introduced a set of 2-forms and a complex $\mathfrak{su}(2)$-valued connection 1-form and was able to describe complex general relativity as a constrained BF theory. It was later realized that by solving the constraint on the 2-forms both the Ashtekar's formulation \cite{Capo} and self-dual Palatini action were recovered \cite{Samuel,Smolin}.

The problems with the reality conditions continued until a different path was taken. Barbero \cite{Barbero} showed that it was possible to introduce real Ashtekar variables instead of complex ones and this approach became very suitable to attack again the problem of quantizing gravity.  Afterwards, Holst introduced a modification of the Hilbert-Palatini action and it was showed that Barbero's formulation was the Hamiltonian form of this action principle \cite{Holst, Barros}. Moreover, real general relativity was also written as a constrained BF theory \cite{Reise,Pietri} and later works generalized the constraint on the B field \cite{Montes1, MontesMer} in such a way that by solving it the Immirzi parameter $\gamma$ \cite{Immi}, a free parameter appearing in Barbero's formulation, was naturally included.

% Some changes concerning the importance of Immirzi parameter in Lorentz-covariant formulation of gravity
% We add the bibliografy \bibitem{Alexandrov3} Alexandrov S 2002 Choice of connection in loop quantum gravity \href{http://prd.aps.org/abstract/PRD/v65/i2/e024011}{\texti{Phys. Rev.} D {\bf 65} 024011}
%Comment on Black hole entropy independence of Immirzi parameter at the end of the paragraph

Although this parameter has a topological nature at the classical level \cite{Perez2,Kaul}, at the quantum level it enters explicitly in the spectra of geometric operators \cite{Lewa,RovSmo} as well as in the expression of the black hole entropy \cite{Rovelli1,ABK,Meis,Agul,JKA}. Nonetheless, its presence is still unclear and its physical relevance is a matter of controversy \cite{Thiem}. In the works \cite{Alexandrov,Alexandrov1,Alexandrov2} it has been shown that in a Lorentz-covariant formulation of general relativity it is possible to eliminate this parameter from the spectrum of the area operator, but the resultant connection is non-commutative and this hinders us to understand the geometric meaning of the theory and to build up its appropriate Hilbert space. The construction of this non-commutative connection (which in fact transforms as a true spacetime connection) is done through the introduction of the Dirac bracket to handle the second-class constraints arising in the Hamiltonian analysis of the Holst action. Note that this connection comes from a general connection parametrized by two numbers. Furthermore, it is possible to choose this numbers to produce a commutative connection (a Lorentz-covariant generalization of Barbero's connection), which leads to the usual $\gamma$-dependent spectrum and to the same Hilbert space of loop quantum gravity, but that fails to transform correctly under time diffeomorphisms \cite{Alexandrov3}. These last results have been confirmed in \cite{Karim,Geiller}, where the authors construct a commutative Lorentz-covariant connection by explicitly solving the second-class constraints. As a final comment, in a recent work a black hole entropy independent of the Immirzi parameter has been found \cite{Bianchi}, which seems to be compatible with a quantum theory of gravity based on a non-commutative connection in the sense discussed above.

%(a recent work has found a black hole entropy independent of the Immirzi parameter \cite{Bianchi})

% Some recent works have proposed that such a parameter should not appear in the quantum theory of gravity as a consequence of the full Lorentz covariance of the theory \cite{Alexandrov,Alexandrov1,Alexandrov2}, a property that Barbero's formulation lacks because a partial gauge-fixing of Lorentz invariance is needed to obtain it. However, this result seems to contradict the results of Refs. \cite{Karim,Geiller}, where making use of an alternative procedure to preserve Lorentz covariance explicitly (solving the second-class constraints instead of using the Dirac brackets), it is shown that the Hilbert space of the Lorentz covariant theory can be reduced to that of the SU(2) theory and thus the spectrum of the area operator in the Lorentz covariant theory matchs exactly the area spectrum of Loop Quantum Gravity. 

%The latter aproach seems more reliable since observable quantities should not depend of any gauge-fixing.

%Since BF gravity as developed in this paper will be fully Lorentz covariant, we expect this work will contribute to this issue

% Modification concerning the spin networks as describing the spatial geometry. I add the reference \bibitem{AshLew} Ashtekar A and Lewandowski J 2004 Background independent quantum gravity: a status report \href{http://iopscience.iop.org/0264-9381/21/15/R01/}{\textit{Class Quantum Grav.} {\bf 21} R53}

In the recent years, a novel way of facing the problem of quantizing the gravitational field has been developed. It is known as the spin foam approach \cite{Perez,Oriti,Baez,Baez1,Engle} and intends a path integral quantization of gravity. The starting point of any spin foam model is a BF-type action principle and since gravity can be formulated in these terms, its path integral quantization can be in principle implemented. The spin foam approach supplies the loop approach, since the latter describes the spatial quantum geometry (the dynamics of spin networks can in principle be generated by promoting the scalar constraint to a quantum operator, but there are ambiguities in the definition of it \cite{AshLew}) but the former can describe the spacetime quantum geometry; in fact, the spin networks of loop quantum gravity can be seen as slices of a spin foam, and then the evolution of spin networks can be better understood in the spin foam formalism. This shows that BF theories are a valuable tool to attack the problem of quantizing gravity and the study of their Hamiltonian structure contributes to that issue. The Lorentz-covariant Hamiltonian analysis presented here supplements the previous results presented in \cite{mitesis,mitesis1}.

%Finally, BF theories of gravity have been shown to be suitable to build the so-called spin foam models of gravity  \cite{Perez,Oriti,Baez,Baez1,Engle}, where a path integral quantization of gravity is intended. Thus BF theories are a valuable tool to attack the problem of quantizing gravity and the study of the Hamiltonian structure of this theories contributes to this issue.

% I add the reference \bibitem{Roche} Buffenoir E, Henneaux M, Noui K and Roche Ph 2004 Hamiltonian analysis of Plebanski Theory \href{http://iopscience.iop.org/0264-9381/21/22/012/}{\textit{Class. Quantum Grav.} {\bf 21} 5203}
% Here I explain a little the difference between Krasnov, Alexandrov and Buffenoir et al.

This paper is organized as follows. In Sect. 2, by the sake of completeness, we develop the (reduced) Hamiltonian analysis of pure BF theory including two topological terms (the Pontrjagin and Euler invariants) and two volume terms, and show that this complete action is of topological character. We then perform the Hamiltonian analysis of BF gravity with Immirzi parameter and two volume terms in Sect. 3. Finally, we perform the canonical analysis of the CMPR action principle for gravity in Sect. 4. As in the case of the Hamiltonian analysis of Holst action, second-class constraints arise in both BF action principles for general relativity. The Lorentz-covariant Hamiltonian analysis of BF-like actions for general relativity (without the Immirzi parameter) has been already performed in \cite{Roche} and \cite{Krasn}; the main difference between the actions considered there is that in the former work the Lagrange multiplier (which imposes the constraint on the B field) carries spacetime indexes, but in the latter it carries Lorentz indexes. The approach of the former then leads to quadratic terms in $\tensor{B}{_{0a}^{IJ}}$ implying we must introduce the canonical momentum conjugated to it and thus complicating the Hamiltonian analysis. To avoid this issue, we follow the approach of the latter, where  $\tensor{B}{_{0a}^{IJ}}$ appears linearly.   

\section{Hamiltonian analysis of BF theory with topological and volume terms}

% Here I introduce the notation \tilde{\eta}^{abc}

We assume that the spacetime manifold $M$ is four-dimensional throughout this paper. We also consider $SO(4)$ or $SO(3,1)$ as the internal gauge groups of our action principles, the internal metric being $\eta_{IJ}=$diag$(\sigma,1,1,1)$ with $\sigma=$-1 (+1) for the SO(3,1) (SO(4)) case (group indexes are raised and lowered using this metric). The Hodge dual is defined by $*T_{IJ}=(1/2)\epsilon_{IJKL}T^{KL}$, where $T_{IJ}=-T_{JI}$ and $\epsilon_{0123}=1$. In the case of spacetime indexes, the tensor density $\tilde{\eta}^{\mu\nu\alpha\beta}$ ($\tilde{\eta}^{0123}=1$) is totally antisymmetric and we define $\tilde{\eta}^{abc}\equiv\tilde{\eta}^{0abc}$. Similarly, $\utilde{\eta}_{abc}$ is totally antisymmetric and such that $\utilde{\eta}_{123}=1$ 

An action principle for pure BF theory is given by

\begin{equation}\label{eq1}
S[A,B]=\int_M B^{IJ}\wedge F_{IJ}[A],
\end{equation}
 
\noindent and it is well known that this action defines a topological field theory (see for instance Ref. \cite{Montes2}). This action is both diffeomorphism and Lorentz invariant, and we want to consider the more general action principle involving the pure BF theory respecting these two symmetries that can be constructed from $B^{IJ}$ and $\tensor{A}{_I^J}$. The action principle considered is \cite{Montes3}

\begin{eqnarray}
\hspace{-10mm}S[A,B]=\int_{M}\Biggl[&\left(B^{IJ}+\frac{1}{\gamma}*B^{IJ}\right)\wedge F_{IJ}+b_1 B^{IJ}\wedge B_{IJ}\nonumber\\
&+b_2 *B^{IJ}\wedge B_{IJ}+\theta_1 \tensor{F}{^I_J}\wedge\tensor{F}{^J_I}+\theta_2*F^{IJ}\wedge F_{IJ}\Biggr],\label{eq2}
\end{eqnarray}

\noindent where we have included two BF terms and the coupling constants $b_1$ and $b_2$ introduce two volume terms. The terms proportional to $\theta_1$ and $\theta_2$ are the Pontrjagin and the Euler classes, respectively. These topological invariants can be rewritten as exterior derivatives of some appropriate 3-forms and the equations of motion do not depend on them; in fact, the equations of motion following from (\ref{eq2}) are

\begin{eqnarray}
F_{IJ}+\frac{1}{\gamma}&*F_{IJ}+2b_1B_{IJ}+2b_2*B_{IJ}=0,\label{eq3}\\
&DB^{IJ}+\frac{1}{\gamma}D*B^{IJ}=0,\label{eq4}
\end{eqnarray}

\noindent which do not depend on $\theta_1$, $\theta_2$ as long as the Bianchi's identity $DF^{IJ}=0$ holds. Nevertheless, the canonical structure of the theory is affected by these topological terms and one might expect some off-shell non-vanishing effects at the quantum level. However, our treatment will be purely classical. The inclusion of topological terms in the context of general relativity has been already considered; see \cite{Montes4,Perez2,Kaul} for details.

To perform the (3+1) decomposition of the action (\ref{eq2}) we assume that the manifold $M$ has a topology $\mathbb{R}\times\Sigma$, where $\Sigma$ is a compact 3-manifold without a boundary ($\partial\Sigma=0$). We then choose a foliation of spacetime in terms of the level hyper-surfaces of a global time function $t$ such that $t=$const has the topology of $\Sigma$. Let $x^a$ ($a=1,2,3$) be the coordinates on $\Sigma$; the action (\ref{eq2}) can be written as

\begin{eqnarray}
 \hspace{-20mm}S[A,B]=\int_{\mathbb{R}}&dt\int_{\Sigma} d^3x\Biggl[\left(\mathop{\Pi}^{(\gamma)}\tensor{}{^a_{IJ}}-\theta_1\tilde{\eta}^{abc}F_{bcIJ}+\theta_2\tilde{\eta}^{abc}*F_{bcIJ}\right)\tensor{\dot{A}}{_a^{IJ}}\nonumber\\
&\hspace{-10mm}+A_{0IJ}D_a\mathop{\Pi}^{(\gamma)}\tensor{}{^{aIJ}}+\tensor{B}{_{0a}^{IJ}}\left(\frac{1}{2}\tilde{\eta}^{abc}\mathop{F}^{(\gamma)}\tensor{}{_{bcIJ}}+2b_1\tensor{\Pi}{^a_{IJ}}+2b_2*\tensor{\Pi}{^a_{IJ}}\right)\Biggr],\label{eq5}
\end{eqnarray}

\noindent where $\Pi^{aIJ}\equiv\frac{1}{2}\tilde{\eta}^{abc}\tensor{B}{_{bc}^{IJ}}$ and the boundary terms have been neglected; we have also introduced the useful notation
\begin{equation}
 \mathop{V}^{(\gamma)}\tensor{}{^{IJ}}\equiv V^{IJ}+\frac{1}{\gamma}*V^{IJ}.
\end{equation}
  
As can be seen from (\ref{eq5}), the topological terms contribute to the conjugate momenta of $A_{aIJ}$ generating canonical transformations. By performing the change of variables $\tensor{P}{^a_{IJ}}\equiv\stackrel{(\gamma)}{\Pi}\tensor{}{^a_{IJ}}-\theta_1\tilde{\eta}^{abc}F_{bcIJ}+\theta_2\tilde{\eta}^{abc}*F_{bcIJ}$, Eq. (\ref{eq5}) takes the form

\begin{equation}\label{eq6}
S[A,P]=\int_{\mathbb{R}}dt\int_{\Sigma} d^3x\left(P^{aIJ}\dot{A}_{aIJ}+A_{0IJ}\mathcal{G}^{IJ}+B_{0aIJ}\Psi^{aIJ}\right),
\end{equation}

\noindent where the tensor densities of weight 1, $\mathcal{G}^{IJ}$ and $\Psi^{aIJ}$, have the following expressions

\begin{eqnarray}
&\hspace{-22mm}\mathcal{G}^{IJ}\equiv D_aP^{aIJ}=\partial_a P^{aIJ}+\tensor{A}{_a^I_K}P^{aKJ}+\tensor{A}{_a^J_K}P^{aIK},\label{eq7}\\
&\hspace{-22mm}\Psi^{aIJ}\equiv \frac{1}{2}\tilde{\eta}^{abc}\mathop{F}^{(\gamma)}\tensor{}{_{bc}^{IJ}}+\frac{2\gamma^2}{\gamma^2-\sigma}\Biggl\{\left(b_1-\frac{\sigma}{\gamma}b_2\right)P^{aIJ}+\left(b_2-\frac{b_1}{\gamma}\right)*P^{aIJ}\nonumber\\
&\hspace{20mm}+\left[b_1\left(\theta_1+\frac{\sigma}{\gamma}\theta_2\right)-\sigma b_2\left(\theta_2+\frac{\theta_1}{\gamma}\right)\right]\tilde{\eta}^{abc}\tensor{F}{_{bc}^{IJ}}\nonumber\\
&\hspace{20mm}+\left[b_2\left(\theta_1+\frac{\sigma}{\gamma}\theta_2\right)-b_1\left(\theta_2+\frac{\theta_1}{\gamma}\right)\right]\tilde{\eta}^{abc}*\tensor{F}{_{bc}^{IJ}}\Biggr\}.\label{eq8}
\end{eqnarray}

Since $A_{0IJ}$ and $B_{0aIJ}$ appear linearly in the action (\ref{eq6}), they play the role of Langrange multipliers and impose the primary constraints $\mathcal{G}^{IJ}\approx 0$ and $\Psi^{aIJ}\approx 0$. The Hamiltonian of the theory is $H=-\int_{\Sigma}d^3x\left(A_{0IJ}\mathcal{G}^{IJ}+B_{0aIJ}\Psi^{aIJ}\right)$ and the canonical pair $(A,P)$ must satisfy the relation

\begin{equation}
\left\{A_{aIJ}(x),P^{bKL}(y)\right\}=\delta_a^b\delta_I^{[K}\delta_J^{L]}\delta^3(x,y).\label{eq9}
\end{equation}

We see that $H$ weakly vanishes as a consequence of the weak vanishing of the constraints; this is something to be expected in any diffeomorphism-invariant theory. Now we need to evolve the constraints in order to know the full content of the constraints of the theory. According to Dirac formalism \cite{Teitel}, the evolution of the system must preserve the constraints and this implies that $\left\{C,H\right\}\approx 0$ for each constraint $C$. The Poisson brackets among the constraints is given by

\begin{eqnarray}
 &\hspace{-25mm}\left\{\mathcal{G}^{IJ}(x),\mathcal{G}^{KL}(y)\right\}=\frac{1}{2}\left(-\eta^{IK}\mathcal{G}^{JL}+\eta^{JK}\mathcal{G}^{IL}+\eta^{IL}\mathcal{G}^{JK}-\eta^{JL}\mathcal{G}^{IK}\right)\delta^3(x,y),\label{eq10}\\
 &\hspace{-25mm}\left\{\Psi^{aIJ}(x),\Psi^{bKL}(y)\right\}=0,\label{eq11}\\
 &\hspace{-25mm}\left\{\Psi^{aIJ}(y),\mathcal{G}^{KL}(y)\right\}=\frac{1}{2}\left(-\eta^{IK}\Psi^{aJL}+\eta^{JK}\Psi^{aIL}+\eta^{IL}\Psi^{aJK}-\eta^{JL}\Psi^{aIK}\right)\delta^3(x,y).\label{eq12}
\end{eqnarray}

Since the constraint algebra closes, the evolution of the constraints is trivial and we find no more constraints. Therefore, the constraints $\mathcal{G}^{IJ}$ and $\Psi^{aIJ}$ are first-class and generate all the gauge symmetries of the theory, but they constitute a reducible set because the Bianchi's identity implies the relation (see also \cite{Montes2,Mondra})

% Here I change the expression of the reduciblity condition for an equivalent expression. The first expression that we considered can be obtained from this one making some algebra, and is valid only when b_1\neq\pm b_2 (in the Euclidean case), since this condition causes a vanishing of the determinant of the matrix transformation from D_a\Psi^a to \mathcal{G}.

\begin{equation}
 \hspace{-7mm}D_a\Psi^{aIJ}-\frac{2\gamma^2}{\gamma^2-\sigma}\left[\left(b_1-\frac{\sigma}{\gamma}b_2\right)\mathcal{G}^{IJ}+\left(b_2-\frac{b_1}{\gamma}\right)*\mathcal{G}^{IJ}\right]=0\label{eq13}
\end{equation}

\noindent This expression amounts to 6 equations relating both constraints and we are left only with 18 independent first-class constraints; there are no second-class constraints. Since there are 18 configuration variables $A_{aIJ}$, the system has $(2\times18-2\times18-0)/2=0$ local degrees of freedom. 

Therefore, adding the volume terms and the two topological invariants considered above to the BF action does not modify its topological character. Despite the complicated constraint (\ref{eq8}), the constraint algebra (\ref{eq10})-(\ref{eq12}) is the same as the constraint algebra of the pure BF action \cite{Montes2}.

\section{Hamiltonian analysis of BF gravity plus cosmological constant}

Now we consider the canonical analysis of real BF gravity with the Immirzi (or the Barbero-Immirzi) parameter coupled to a cosmological constant. In the context of BF theories, this coupling has been successfully accomplished by considering the following action principle \cite{MontesMer2} (see \cite{MontesMer3} for another alternative approach)

\begin{eqnarray}
 &\hspace{-24mm}S[B,A,\phi,\mu]=\int_{M}\Biggl[\left(B^{IJ}+\frac{1}{\gamma}*B^{IJ}\right)\wedge F_{IJ}-\phi_{IJKL}B^{IJ}\wedge B^{KL}-\mu\phi_{IJKL}\epsilon^{IJKL}\nonumber\\
 &\hspace{15mm}+\mu\lambda+l_1B_{IJ}\wedge B^{IJ}+l_2 B_{IJ}\wedge *B^{IJ}\Biggr].\label{eq14}
\end{eqnarray}

Here $\phi_{IJKL}$ is an internal tensor with the index symmetries $\phi_{IJKL}=\phi_{KLIJ}$, $\phi_{IJKL}=-\phi_{JIKL}$ and $\phi_{IJKL}=-\phi_{IJLK}$; the 4-form $\mu$ ($\mu_{0123}\equiv\mu_0$) implies the additional restriction $\phi_{IJKL}\epsilon^{IJKL}=\lambda$ on $\phi_{IJKL}$; $\lambda$, $l_1$ and $l_2$ are just constants related to the cosmological constant.

By performing the (3+1) decomposition of the action (\ref{eq14}), we obtain

\begin{eqnarray}
 &\hspace{-23mm}S[A,B,\phi,\mu]=\int_{\mathbb{R}}dt\int_{\Omega}d^3x\Biggl[\mathop{\Pi}^{(\gamma)}\tensor{}{^{aIJ}}\dot{A}_{aIJ}+A_{0IJ}D_a\mathop{\Pi}^{(\gamma)}\tensor{}{^{aIJ}}+\frac{1}{2}\tilde{\eta}^{abc}\mathop{F}^{(\gamma)}\tensor{}{_{abIJ}}\tensor{B}{_{0c}^{IJ}}\nonumber\\
 &\hspace{-15mm}-\left(2\tensor{B}{_{0a}^{IJ}}\Pi^{aKL}+\mu_0\epsilon^{IJKL}\right)\phi_{IJKL}+\mu_0\lambda+2l_1\Pi^{aIJ}B_{0aIJ}+2l_2*\Pi^{aIJ}B_{0aIJ}\Biggr],\label{eq15}
\end{eqnarray}

\noindent with $\Pi^{aIJ}$ defined as in Sect. 2. Now we shall use the equation of motion correspondig to $\phi_{IJKL}$ to put the components $B_{0aIJ}$ in terms of $\Pi^{aIJ}$, thus eliminating this internal tensor from the action principle; this procedure is essentially the one followed to get the Hamiltonian formulation of the Pleba\'{n}ski action \cite{Capo} (see also \cite{MontesMer4} and \cite{Krasn}). The equation of motion for $\phi_{IJKL}$ is 

%In order to avoid the introduction of too many variables (which complicates the Hamiltonian analysis), we shall use the equation of motion correspondig to $\phi_{IJKL}$ to put the components $B_{0aIJ}$ in terms of $\Pi^{aIJ}$, thus eliminating this internal tensor from the action principle; this procedure is essentially the one followed to get the Hamiltonian formulation of the Pleba\'{n}ski action \cite{Capo} (see also \cite{MontesMer4} and \cite{Krasn}). The equation of motion for $\phi_{IJKL}$ is

\begin{equation}
 \tensor{B}{_{0a}^{IJ}}\Pi^{aKL}+\tensor{B}{_{0a}^{KL}}\Pi^{aIJ}+\mu_0\epsilon^{IJKL}=0\label{eq16},
\end{equation}

\noindent that implies $\mu_0=-\sigma \mathcal{V}/4$, where
\begin{equation}
 \mathcal{V}\equiv\frac{1}{24}\tilde{\eta}^{\mu\nu\lambda\rho}\epsilon_{IJKL}\tensor{B}{_{\mu\nu}^{IJ}}\tensor{B}{_{\lambda\rho}^{KL}}=\frac{1}{3}\epsilon_{IJKL}\tensor{B}{_{0a}^{IJ}}\Pi^{aKL}\label{eq17}
\end{equation}

\noindent is the four-dimensional volume; we shall assume that it does not vanish anywhere.

To handle the expression (\ref{eq17}), we shall introduce the following quantities:

\begin{eqnarray}
 & N^a\equiv\frac{\sigma}{2h}\tilde{\eta}^{abc}h_{bd}\tensor{B}{_{0c}^{IJ}}\tensor{\Pi}{^d_{IJ}}, \hspace{10mm}N\equiv\frac{\mathcal{V}}{h}\label{var1}\\
 & hh^{ab}\equiv\frac{\sigma}{2}\Pi^{aIJ}\tensor{\Pi}{^b_{IJ}},\hspace{10mm} \Phi^{ab}\equiv-\sigma *\tensor{\Pi}{^a_{IJ}}\Pi^{bIJ},\label{var2}
 \end{eqnarray}

% We modify the order of equations and make explicit the dependence on \stackrel{(\gamma)}{\Pi} in \Phi^{ab} and hh^{ab}

\noindent where $(h_{ab})$ is the inverse of $(h^{ab})$, $h=\det(h_{ab})$ and $N\neq0$. In terms of the $\gamma$-valued variable $\stackrel{(\gamma)}{\Pi}$, $\Phi^{ab}$ and $hh^{ab}$ take the following form

\begin{eqnarray}
 &hh^{ab}={\textstyle\eta\left[\stackrel{(\gamma)}{(hh^{ab})}+\frac{\gamma^{-1}}{1+\sigma\gamma^{-2}}\stackrel{(\gamma)}{\Phi}\tensor{}{^{ab}}\right]}\label{var3},\\
 &\Phi^{ab}{\textstyle=\eta\left[\stackrel{(\gamma)}{\Phi}\tensor{}{^{ab}}+\frac{4\sigma\gamma^{-1}}{1+\sigma\gamma^{-2}}\stackrel{(\gamma)}{(hh^{ab})}\right]},\label{var4}
\end{eqnarray}

\noindent where $\stackrel{(\gamma)}{(hh^{ab})}$ and $\stackrel{(\gamma)}{\Phi}\tensor{}{^{ab}}$ are the expressions (\ref{var2}) with the replacement $\Pi\rightarrow\stackrel{(\gamma)}{\Pi}$ and $\eta\equiv\frac{\gamma^2(\gamma^2+\sigma)}{(\gamma^2-\sigma)^2}$.

After some algebra we can show that $\tensor{B}{_{0a}^{IJ}}$ can be expressed in terms of the introduced variables (\ref{var1}) and (\ref{var2})  as

\begin{equation}
 \hspace{-5mm}\tensor{B}{_{0a}^{IJ}}=\frac{1}{8}Nh_{ab}\epsilon^{IJKL}\tensor{\Pi}{^b_{KL}}+\frac{1}{2}\utilde{\eta}_{abc}\Pi^{bIJ}N^{c}+\frac{1}{16h}Nh_{ac}h_{bd}\Pi^{bIJ}\Phi^{cd}.\label{eq18}
\end{equation}

By substituting this expression into (\ref{eq16}) and taking the appropriate contractions, and since $N\neq0$, we find that $\Phi^{ab}$ must satisfy

\begin{equation}
 \Phi^{ab}\left(h^2\delta^c_b-\frac{\sigma}{4}h_{bd}h_{fe}\Phi^{de}\Phi^{fc}\right)=0.
\end{equation}

% At the end of this paragraph, I add something could answer why the other two answers are not allowed.

This equation has the solutions $\Phi^{ab}=0$ and $\Phi^{ab}=\pm 2\sqrt{\sigma}hh^{ab}$. In the Lorentzian case, $\Phi^{ab}=0$ is the only real solution. Because we are interested in real variables, this will be the only one considered below. In fact, the solutions $\Phi^{ab}=\pm 2\sqrt{\sigma}hh^{ab}$ are not allowed in both signatures since they are incompatible with the solution for the 2-forms $B^{IJ}$ in terms of the tetrad field \footnote[1]{If the constraint on the B fields involves only the Lorentz invariant $\epsilon_{IJKL}\phi^{IJKL}$, then $B^{IJ}\sim *e^I\wedge e^J$ or $B^{IJ}\sim e^I\wedge e^J$. Using these expressions, the equation $\Phi^{ab}\propto hh^{ab}$ cannot be satisfied.}.

In this way, the 20 Eqs. (\ref{eq16}) (taking into account the value of $\mu_0$) plus the 4 expressions defining $N$ and $N^a$ (which remain unknown) yield to the expression (\ref{eq18}) for the 18 components $\tensor{B}{_{0a}^{IJ}}$ and the 6 conditions on $\Phi^{ab}$.

By substituting Eq. ($\ref{eq18}$) into Eq. ($\ref{eq15}$), the action becomes

\begin{eqnarray}
 &\hspace{-20mm}S[A,\Pi]=\int_{\mathbb{R}}dt\int_{\Omega}d^3x \Biggl[\mathop{\Pi}^{(\gamma)}\tensor{}{^{aIJ}}\dot{A}_{aIJ}+A_{0IJ}D_a\mathop{\Pi}^{(\gamma)}\tensor{}{^{aIJ}}+\frac{1}{2}\mathop{\Pi}^{(\gamma)}\tensor{}{^{bIJ}}F_{baIJ}N^a\nonumber\\
 &+\left(\frac{1}{8}\tilde{\eta}^{abc}h_{ad}*\mathop{\Pi}^{(\gamma)}\tensor{}{^{dIJ}}\tensor{F}{_{bcIJ}}+\Lambda h\right)N+\lambda_{ab}\Phi^{ab}\Biggr],\label{eq19}
\end{eqnarray}
  
\noindent where $\Lambda\equiv3l_2-\sigma\lambda/4$ and $\lambda_{ab}$ is a Lagrange multiplier enforcing the constraint $\Phi^{ab}=0$. Since $A_{0IJ}$, $N$ and $N^a$ appear linearly in the action, they play the role of Lagrange multipliers and impose, with $\lambda_{ab}$, the primary constraints

\begin{eqnarray}
 &\hspace{-13mm}\mathcal{G}^{IJ}\equiv D_a\mathop{\Pi}^{(\gamma)}\tensor{}{^{aIJ}}\approx 0,\hspace{10mm}(\textrm{Gauss constraint});\label{eq20}\\
 &\hspace{-13mm}\mathcal{H}\equiv\frac{1}{8}\tilde{\eta}^{abc}h_{ad}*\mathop{\Pi}^{(\gamma)}\tensor{}{^{dIJ}} F_{bcIJ}+\Lambda h\approx0,\hspace{10mm}(\textrm{scalar constraint});\label{eq21}\\
 &\hspace{-13mm}\mathcal{H}_a\equiv\frac{1}{2}\mathop{\Pi}^{(\gamma)}\tensor{}{^{bIJ}}F_{baIJ}\approx 0,\hspace{10mm} (\textrm{vector constraint});\label{eq22}\\
 &\hspace{-13mm}\Phi^{ab}\approx 0.\label{eq23}
\end{eqnarray}

The primary Hamiltonian is then given by
\begin{equation}\label{eq24}
 H=-\int_{\Omega}d^3x\left(A_{0IJ}\mathcal{G}^{IJ}+N\mathcal{H}+N^a\mathcal{H}_a+\lambda_{ab}\Phi^{ab}\right),
\end{equation}

\noindent which weakly vanishes as a consequence of the constraints. The canonical pair $(A,\stackrel{(\gamma)}{\Pi})$, which parametrizes the phase space of the theory, satisfies a relation similar to Eq. (\ref{eq9}).

% We modify the constraint algebra to put the main part of the Poisson brackets in terms of \stackrel{(\gamma)}{\Pi} 

The primary constraint algebra is

\begin{eqnarray}
 &\hspace{-25mm}\left\{\mathcal{G}^{IJ}(x),\mathcal{G}^{KL}(y)\right\}=\frac{1}{2}\left(-\eta^{IK}\mathcal{G}^{JL}+\eta^{JK}\mathcal{G}^{IL}+\eta^{IL}\mathcal{G}^{JK}-\eta^{JL}\mathcal{G}^{IK}\right)\delta^3_{x,y},\label{eq25}\\
 &\hspace{-25mm}\left\{\mathcal{G}^{IJ}(x),\mathcal{H}_a(y)\right\}=0,\label{eq26}\\
 &\hspace{-25mm}\left\{\mathcal{G}^{IJ}(x),\mathcal{H}(y)\right\}=0,\label{eq27}\\
 &\hspace{-25mm}\left\{\mathcal{G}^{IJ}(x),\Phi^{ab}(y)\right\}=0,\label{eq28}\\
 &\hspace{-25mm}\left\{\mathcal{H}_a(x),\mathcal{H}_b(y)\right\}=\textstyle{\left[\frac{1}{2}\mathcal{H}_a(y)\frac{\partial}{\partial y^b}-\frac{1}{2}\mathcal{H}_b(x)\frac{\partial}{\partial x^a}-\frac{1}{4}F_{abIJ}\mathcal{G}^{IJ}\right]\delta^3_{x,y}},\label{eq29}\\
 &\hspace{-25mm}\left\{\mathcal{H}_a(x),\mathcal{H}(y)\right\}=\textstyle{\left[\frac{1}{2}\mathcal{H}(y)\frac{\partial}{\partial y^a}-\frac{1}{2}\mathcal{H}(x)\frac{\partial}{\partial x^a}-\frac{\sigma\Lambda\eta}{4}h_{ab}\left(\stackrel{(\gamma)}{\Pi}\tensor{}{^b_{IJ}}-\frac{2\gamma^{-1}}{1+\sigma\gamma^{-2}}*\stackrel{(\gamma)}{\Pi}\tensor{}{^b_{IJ}}\right)\mathcal{G}^{IJ}\right]\delta^3_{x,y}}\nonumber\\
 &\hspace{-21mm}-{\textstyle\frac{1}{32}\tilde{\eta}^{cde}F_{deKL}\left[\tensor{\epsilon}{_{IJ}^{KL}}h_{ac}+\frac{\sigma\eta}{h}*\stackrel{(\gamma)}{\Pi}\tensor{}{^{fKL}}\left(\stackrel{(\gamma)}{\Pi}\tensor{}{^r_{IJ}}-\frac{2\gamma^{-1}}{1+\sigma\gamma^{-2}}*\stackrel{(\gamma)}{\Pi}\tensor{}{^r_{IJ}}\right)H_{arcf}\right]\mathcal{G}^{IJ}\delta^3_{x,y}},\label{eq30}\\
 &\hspace{-25mm}\left\{\mathcal{H}_a(x),\Phi^{bc}(y)\right\}={\textstyle\left[\frac{1}{2}\Phi^{bc}(y)\frac{\partial}{\partial y^a}-\frac{1}{2}\Phi^{bc}(x)\frac{\partial}{\partial x^a}-\delta_a^{(b}\Phi^{c)d}(y)\frac{\partial}{\partial y^d}\right]\delta^3_{x,y}}\nonumber\\
 &\hspace{10mm}{\textstyle+\sigma\eta\delta_a^{(b}\left(*\stackrel{(\gamma)}{\Pi}\tensor{}{^{c)}_{IJ}}-\frac{2\sigma\gamma^{-1}}{1+\sigma\gamma^{-2}}\stackrel{(\gamma)}{\Pi}\tensor{}{^{c)}_{IJ}}\right)\mathcal{G}^{IJ}\delta^3_{x,y}},\label{eq31}\\
 &\hspace{-25mm}\left\{\mathcal{H}(x),\mathcal{H}(y)\right\}={\textstyle\frac{\sigma}{8}\left[h(x)h^{ab}(x)\mathcal{H}_a(x)\frac{\partial}{\partial x^b}-h(y)h^{ab}(y)\mathcal{H}_a(y)\frac{\partial}{\partial y^b}\right]\delta^3_{x,y}}\nonumber\\
 &\hspace{-22mm}{\textstyle+\frac{1}{8}\left[\frac{1}{4}\tilde{\eta}^{abe}*\stackrel{(\gamma)}{\Pi}\tensor{}{^{gIJ}}F_{beIJ}h^{mc}\utilde{\eta}_{rm(a}h_{g)s}+\Lambda\tilde{\eta}^{abc}h_{ar}h_{bs}\right](x)\Phi^{rs}(x)\frac{\partial}{\partial x^c}\delta^3_{x,y}}-(x\leftrightarrow y),\label{eq32}\\
 &\hspace{-25mm}\left\{\mathcal{H}(x),\Phi^{ab}(y)\right\}={\textstyle-\frac{1}{4}\Psi^{ab}\delta^3_{x,y}},\label{eq33}\\
 &\hspace{-25mm}\left\{\Phi^{ab}(x),\Phi^{cd}(y)\right\}=0\label{eq34},
\end{eqnarray}

\noindent where $\delta^3_{x,y}$ is a shorthand for Dirac's delta, $H_{abcd}\equiv h_{ab}h_{cd}-h_{ac}h_{bd}-h_{ad}h_{bc}$ and $\Psi^{ab}$ has the following expression
\begin{equation}\label{eq35}
 \Psi^{ab}\equiv -2\eta h_{cf}\left(-\stackrel{(\gamma)}{\Pi}\tensor{}{^f_{IJ}}+\frac{2\gamma^{-1}}{1+\sigma\gamma^{-2}}*\stackrel{(\gamma)}{\Pi}\tensor{}{^f_{IJ}}\right)\tilde{\eta}^{(a|cd}D_d\stackrel{(\gamma)}{\Pi}\tensor{}{^{|b)IJ}}.
\end{equation}
 
\noindent We see that the constraint algebra fails to close due to the Poisson bracket (\ref{eq33}), which implies that secondary constraints could arise in the theory. Thus, whereas the evolution of the constraints $\mathcal{G}^{IJ}$ and $\mathcal{H}_a$ leads to $0=0$, giving no more constraints or conditions on the Langrange multipliers, the evolution of the constraint $\Phi^{ab}$ leads to the condition

\begin{equation}
 N\Psi^{ab}\approx 0,
\end{equation}

\noindent whose solution is 

\begin{equation}\label{eq36}
 \Psi^{ab}\approx 0,
\end{equation}

\noindent since $N\neq 0$ by assumption. Therefore, the last expression must be incorporated as a secondary constraint of the theory and thus, using (\ref{eq36}), the evolution of $\mathcal{H}$ is trivially satisfied.

Following Dirac's method, we now need to evolve the constraint $\Psi^{ab}$. We will not calculate all the Poisson brackets involving the constraint $\Psi^{ab}$, but only the required ones to complete the analysis. The Poisson brackets are

\begin{eqnarray}
&\hspace{-25mm}\left\{\Psi^{ab}(x),\mathcal{G}^{IJ}(y)\right\}=0,\label{eq37}\\
&\hspace{-25mm}\left\{\Psi^{ab}(x),\mathcal{H}_c(y)\right\}={\textstyle\left[\frac{1}{2}\Psi^{ab}(y)\frac{\partial}{\partial y^c}-\frac{1}{2}\Psi^{ab}(x)\frac{\partial}{\partial x^c}+\delta^{(a}_c\Psi^{b)d}(x)\frac{\partial}{\partial x^d}\right]\delta^3_{x,y}}\nonumber\\
&\hspace{-21mm}+{\textstyle \frac{\eta}{2}\Biggl[\frac{\sigma\eta}{h}H_{cndf}\left(\stackrel{(\gamma)}{\Pi}\tensor{}{^n_{IJ}}-\frac{2\gamma^{-1}}{1+\sigma\gamma^{-2}}*\stackrel{(\gamma)}{\Pi}\tensor{}{^n_{IJ}}\right)\left(\stackrel{(\gamma)}{\Pi}\tensor{}{^f_{KL}}-\frac{2\gamma^{-1}}{1+\sigma\gamma^{-2}}*\stackrel{(\gamma)}{\Pi}\tensor{}{^f_{KL}}\right)\tilde{\eta}^{(a|de}D_e\stackrel{(\gamma)}{\Pi}\tensor{}{^{|b)KL}}}\nonumber\\
&\hspace{-10mm}{\textstyle+2h_{dc}\tilde{\eta}^{(a|de}\left(D_e\stackrel{(\gamma)}{\Pi}\tensor{}{^{|b)}_{IJ}}-\frac{2\gamma^{-1}}{1+\sigma\gamma^{-2}}*D_e\stackrel{(\gamma)}{\Pi}\tensor{}{^{|b)}_{IJ}}\right)\Biggr]\mathcal{G}^{IJ}\delta^3_{x,y}}\nonumber\\
&\hspace{-21mm}-{\textstyle \eta\delta^{(a}_c\tilde{\eta}^{b)de}h_{df}(x)\left(\stackrel{(\gamma)}{\Pi}\tensor{}{^{f}_{IJ}}(x)-\frac{2\gamma^{-1}}{1+\sigma\gamma^{-2}}*\stackrel{(\gamma)}{\Pi}\tensor{}{^{f}_{IJ}}(x)\right)}(D_x)_e(\mathcal{G}^{IJ}\delta^3_{x,y}),\label{eq38}\\
&\hspace{-25mm}\left\{\Psi^{ab}(x),\Phi^{cd}(y)\right\}=M^{(ab)(cd)}\delta^3_{x,y},\label{eq39}
\end{eqnarray}

\noindent where

\begin{eqnarray}
 &\hspace{-20mm}M^{(ab)(cd)}\equiv{\textstyle 4\sigma\eta^2 h_{ef}\left(\stackrel{(\gamma)}{\Pi}\tensor{}{^f_{IK}}-\frac{2\gamma^{-1}}{1+\sigma\gamma^{-2}}*\stackrel{(\gamma)}{\Pi}\tensor{}{^f_{IK}}\right)}\nonumber\\
 &\hspace{20mm}\times{\textstyle\left[\left(*\stackrel{(\gamma)}{\Pi}\tensor{}{^c_J^K}-\frac{2\sigma\gamma^{-1}}{1+\sigma\gamma^{-2}}\stackrel{(\gamma)}{\Pi}\tensor{}{^c_J^K}\right)\tilde{\eta}^{(a|de}\stackrel{(\gamma)}{\Pi}\tensor{}{^{|b)IJ}}+(c\leftrightarrow d)\right]}\label{matrix1}
\end{eqnarray}

\noindent defines a $6\times 6$ non-singular matrix. Note that the Poisson bracket (\ref{eq38}) is weakly zero. By employing the Poisson brackets (\ref{eq37})-(\ref{eq39}), the evolution of the constraint $\Psi^{ab}$ leads to

\begin{equation}\label{eq40}
 \int_{\Omega}d^3y\left[N(y)\left\{\Psi^{ab}(x),\mathcal{H}(y)\right\}+\lambda_{cd}M^{(ab)(cd)}\delta^3(x,y)\right]\approx 0.
\end{equation}

We need an expression for the Poisson bracket $\left\{\Psi^{ab}(x),\mathcal{H}(y)\right\}$. Using the Jacobi identity

\begin{equation}\label{eq41}
 \hspace{-23mm}{\textstyle\{\{\mathcal{H}(x),\Phi^{ab}(y)\},\mathcal{H}(z)\}=-\{\{\Phi^{ab}(y),\mathcal{H}(z)\},\mathcal{H}(x)\}-\{\{\mathcal{H}(z),\mathcal{H}(x)\},\Phi^{ab}(y)\}}
\end{equation}

\noindent and the fact that $\{\{\mathcal{H}(z),\mathcal{H}(x)\},\Phi^{ab}(y)\}\approx 0$ as a consequence of (\ref{eq32}), (\ref{eq31}) and (\ref{eq34}), we find that

\begin{equation}
 \{\Psi^{ab}(x),\mathcal{H}(z)\}\delta^3(x,y)\approx\{\Psi^{ab}(y),\mathcal{H}(x)\}\delta^3(z,y),
\end{equation}

% I add some lines that maybe explain better where Eq. (51) come from

\noindent where (\ref{eq33}) has been used. Integrating both sides with respect to $z$, we finally obtain

\begin{equation}\label{eq42}
 \{\Psi^{ab}(y),\mathcal{H}(x)\}\approx F^{ab}\delta^3(y,x),
\end{equation}

\noindent with $F^{ab}$ defined by $F^{ab}(x)\equiv\int_{\Omega}d^3z\{\Psi^{ab}(x),\mathcal{H}(z)\}$. By substituting (\ref{eq42}) into (\ref{eq40}), we find

\begin{equation}\label{eq43}
 \lambda_{ab}\approx -\frac{1}{4}N(M^{-1})_{(ab)(cd)}F^{cd}.
\end{equation}

\noindent This implies that all the Langrange multipliers $\lambda_{ab}$ can be fixed; Dirac's method ends here. Now we can count the degrees of freedom of our theory. Substituting Eq. (\ref{eq43}) into Eq. (\ref{eq19}), the Hamiltonian becomes 

\begin{equation}\label{eq44}
 \hspace{-20mm}H=-\int_{\Omega}d^3x\left(A_{0IJ}\mathcal{G}^{IJ}+N\bar{\mathcal{H}}+N^a\mathcal{H}_a\right),\hspace{5mm} \bar{\mathcal{H}}\equiv\mathcal{H}-\frac{1}{4}(M^{-1})_{(ab)(cd)}F^{cd}\Phi^{ab}.
\end{equation}

We shall replace the constraint $\mathcal{H}$ by $\bar{\mathcal{H}}$ since the latter is first-class, as can be readily verified. The constraint algebra listed above shows that $\mathcal{G}^{IJ}$ and $\mathcal{H}_a$ are also first-class and correspond to the generators of local Lorentz transformations and spatial diffeomorphisms, respectively. On the other hand, by virtue of Eq. (\ref{eq39}), $\Phi^{ab}$ and $\Psi^{ab}$ are second-class. Since we have 18 configuration variables $A_{aIJ}$, the number of physical degrees of freedom is $[2\times18-2\times(1+3+6)-(6+6)]/2=2$, the same number as Einstein's theory of gravity (coupled to a cosmological constant).

\section{Hamiltonian analysis of the CMPR action}

The first BF principle to describe pure gravity including the Immirzi parameter was introduced in Ref. \cite{Montes1} and is given by

\begin{equation}\label{eq45}
 \hspace{-25mm}S[B,A,\phi,\mu]=\int_{M}\left[B^{IJ}\wedge F_{IJ}-\phi_{IJKL}B^{IJ}\wedge B^{KL}+\mu\left(a_1\tensor{\phi}{_{IJ}^{IJ}}+a_2\epsilon_{IJKL}\phi^{IJKL}\right)\right].
\end{equation}

\noindent The internal tensor $\phi_{IJKL}$ satisfies the same index symmetries as above, and $a_1$ and $a_2$ are arbitrary constants. As we mentioned, one relevance of this formulation is that the Immirzi parameter is naturally contained within it, since the Holst action is recovered once the constraint on the $B$ fields is solved. The relation between this action and the action principle (\ref{eq14}) was analized at the Lagrangian level in Ref. \cite{MontesMer}.

To perform the Hamiltonian analysis of (\ref{eq45}), we will follow a similar procedure to the one used in Sect. 3. We will not give here all the calculations involved, but the main steps. After the (3+1) decomposition, the equation of motion corresponding to $\phi_{IJKL}$ is

\begin{equation}\label{eq46}
 \tensor{B}{_{0a}^{IJ}}\Pi^{aKL}+\tensor{B}{_{0a}^{KL}}\Pi^{aIJ}-\mu_0\left[a_1\eta^{I[K|}\eta^{J|L]}+a_2\epsilon^{IJKL}\right]=0,
\end{equation}

\noindent which implies $\mu_0=\sigma\mathcal{V}/4a_2$, where the volume $\mathcal{V}$ was defined in (\ref{eq17}). If we now introduce the quantities $h^{ab}$ ($h_{ab}$), $N^a$, $N$($\neq$ 0) and $\Phi^{ab}$ that were introduced in Sect. 3 (Eqs. (\ref{var1}) and (\ref{var2})), the components $\tensor{B}{_{0a}^{IJ}}$ can be expressed in terms of them as

\begin{equation}\label{eq47}
 \hspace{-22mm}\tensor{B}{_{0a}^{IJ}}=\frac{1}{8}Nh_{ab}\epsilon^{IJKL}\tensor{\Pi}{^b_{KL}}+\frac{1}{2}\utilde{\eta}_{abc}\Pi^{bIJ}N^{c}+\frac{1}{16h}Nh_{ac}h_{bd}\Pi^{bIJ}\left(\Phi^{cd}+\frac{a_1}{a_2}hh^{cd}\right).
\end{equation}

% We removed the comparison of this expression with (\ref{eq18})

%in fact, this term will be the main difference between this formulation and the formulation of Sect. 3 since the constraint algebra will be pretty similar.

\noindent Note that this expression differs from (\ref{eq18}) only by the last term. By substituting (\ref{eq47}) into (\ref{eq46}) and after some algebra we find that $\Phi^{ab}$ satisfies the following equation:

\begin{equation}
 \Phi^{ac}\left(2h\delta_c^b-\frac{\sigma}{2h}h_{cd}h_{ef}\Phi^{df}\Phi^{eb}-\frac{\sigma a_1}{2a_2}h_{cd}\Phi^{db}\right)-\frac{2a_1}{a_2}h^2h^{ab}=0,
\end{equation}

\noindent which has the solutions $\Phi^{ab}=-\frac{a_1}{a_2}hh^{ab}$ and $\Phi^{ab}=\pm 2\sqrt{\sigma}hh^{ab}$. As in Sect. 3, we have one real and two complex solutions in the Lorentzian case. In the Euclidean case, all these solutions are real and compatible with the solution for the B's in terms of the tetrad field \footnote[2]{For the constraint considered in the action principle (\ref{eq45}), the solution for the field B is $B^{IJ}=\alpha *e^I\wedge e^J+\beta e^I\wedge e^J$, where $\alpha$ and $\beta$ are constants related to $a_1$ and $a_2$.}, but only the solution $\Phi^{ab}=-\frac{a_1}{a_2}hh^{ab}$ allows an arbitrary Immirzi parameter, whereas the other two solutions fix it to a particular value. We consider below the real solution $\Phi^{ab}=-\frac{a_1}{a_2}hh^{ab}$ (also valid in the Lorentzian signature) and the other two cases $\Phi^{ab}=\pm 2\sqrt{\sigma}hh^{ab}$ can be handled (even in the Lorentzian case) in a similar way by suitably changing the factor $a_1/a_2$.

By using the expression (\ref{eq47}) and neglecting surface terms, the action (\ref{eq45}) becomes

\begin{equation}
\hspace{-18mm}S[A,\Pi]=\int_{\mathcal{R}}dt\int_{\Omega}d^3x\left[\Pi^{aIJ}\dot{A}_{aIJ}+A_{0IJ}\mathcal{G}^{IJ}+N\mathcal{H}+N^a\mathcal{H}_a+\lambda_{ab}\varphi^{ab}\right], 
\end{equation}

\noindent where

\begin{eqnarray}
 &\mathcal{G}^{IJ}\equiv D_a\Pi^{aIJ},\label{cons1}\\
 &\mathcal{H}_a\equiv\frac{1}{2}\Pi^{bIJ}F_{baIJ},\label{cons2}\\
 &\mathcal{H}\equiv\frac{1}{8}\tilde{\eta}^{abc}h_{ad}*\Pi^{dIJ}F_{bcIJ},\label{cons3}\\
 &\varphi^{ab}\equiv\Phi^{ab}+\frac{a_1}{a_2}hh^{ab}.\label{cons4}
\end{eqnarray}

% Here I add a comment relating the these primary constraints to the constraints of section 3

\noindent We have introduced the multipliers $\lambda_{ab}$ to enforce the constraint $\varphi^{ab}=0$, which we obtained a couple of lines above. The primary constraints of the theory are then given by $\mathcal{G}^{IJ}\approx0$, $\mathcal{H}\approx0$, $\mathcal{H}_a\approx0$ and $\varphi^{ab}\approx0$ and the Hamiltonian $H=-\int_{\Omega}d^3x\left(A_{0IJ}\mathcal{G}^{IJ}+N\mathcal{H}+N^a\mathcal{H}_a+\lambda_{ab}\varphi^{ab}\right)$ weakly vanishes as a consequence of them. Note que the number of primary constraints is the same as that in Sect. 3 and that they have a similar structure; in fact, by performing the change $\stackrel{(\gamma)}{\Pi}\rightarrow\Pi$ in (\ref{eq20}) and (\ref{eq22}) we see that they coincide with the constraints (\ref{cons1}) and (\ref{cons2}), respectively; something similar occurs to the constraint (\ref{eq23}), which coincides with the constraint (\ref{cons4}) when we perform the just mentioned change and identify $4\sigma\gamma^{-1}/(1+\sigma\gamma^{-2})$ with $a_1/a_2$. On the other hand, when we make this change in the constraint (\ref{eq21}), the resulting expression is very similar to the corresponding constraint (\ref{cons3}), but because of (\ref{var3}) the factor $h_{ad}\ (\stackrel{(\gamma)}{\Pi}\rightarrow\Pi)$ appearing in (\ref{eq21}) is not equal to the factor $h_{ad}$ of (\ref{cons3}); it is due to this little difference that the constraint algebra (off-shell) of the CMPR action will not be exactly the same as the Sect. 3 when we perform the above change on the canonical momentum.

%The number of constraints is the same as the previous section, but whereas the Immirzi parameter is carried there by the momentum canonically conjugate to $A_{aIJ}$, this information is carried here by the constraint $\varphi^{ab}$.

The algebra of primary constraints is given by

\begin{eqnarray}
&\hspace{-25mm}\{\mathcal{G}^{IJ}(x),\mathcal{G}^{KL}(y)\}=\frac{1}{2}\left(-\eta^{IK}\mathcal{G}^{JL}+\eta^{JK}\mathcal{G}^{IL}+\eta^{IL}\mathcal{G}^{JK}-\eta^{JL}\mathcal{G}^{IK}\right)\delta^3_{x,y}, \label{eq49}\\
&\hspace{-25mm}\{\mathcal{G}^{IJ}(x),\mathcal{C}(y)\}=0\hspace{5mm}(\mathcal{C}=\mathcal{H},\mathcal{H}_a,\varphi^{ab}),\label{eq50}\\
&\hspace{-25mm}\{\mathcal{H}_a(x),\mathcal{H}_b(y)\}={\textstyle \left[\frac{1}{2}\mathcal{H}_a(y)\frac{\partial}{\partial y^b}-\frac{1}{2}\mathcal{H}_b(x)\frac{\partial}{\partial x^a}-\frac{1}{4} F_{abIJ}\mathcal{G}^{IJ}\right]}\delta^3_{x,y},\label{eq51}\\
&\hspace{-25mm}\{\mathcal{H}_a(x),\mathcal{H}(y)\}={\textstyle\frac{1}{2}\left[\mathcal{H}(y)\frac{\partial}{\partial y^a}-\mathcal{H}(x)\frac{\partial}{\partial x^a}\right]\delta^3_{x,y}}\nonumber\\
&\hspace{7mm}{\textstyle -\frac{1}{32}\tilde{\eta}^{cde}F_{deKL}\left[ \tensor{\epsilon}{_{IJ}^{KL}}h_{ac}+\frac{\sigma}{h}*\Pi^{fKL}\tensor{\Pi}{^r_{IJ}}H_{arcf}\right]\mathcal{G}^{IJ}\delta^3_{x,y}},\label{eq52}\\
&\hspace{-25mm}\{\mathcal{H}_a(x),\varphi^{bc}(y)\}={\textstyle\left[\frac{1}{2}\varphi^{bc}(y)\frac{\partial}{\partial y^a}-\frac{1}{2}\varphi^{bc}(x)\frac{\partial}{\partial x^a}-\delta_a^{(b}\varphi^{c)d}(y)\frac{\partial}{\partial y^d}\right]\delta^3_{x,y}}\nonumber\\
&\hspace{10mm}{\textstyle +\sigma\delta_a^{(b}\left(*\tensor{\Pi}{^{c)}_{IJ}}-\frac{a_1}{2a_2}\tensor{\Pi}{^{c)}_{IJ}}\right)\mathcal{G}^{IJ}\delta^3_{x,y}},\label{eq53}\\
&\hspace{-25mm}\{\mathcal{H}(x),\mathcal{H}(y)\}={\textstyle\frac{\sigma}{8}\left[h(x)h^{ab}(x)\mathcal{H}_a(x)\frac{\partial}{\partial x^b}-h(y)h^{ab}(y)\mathcal{H}_a(y)\frac{\partial}{\partial y^b}\right]\delta^3_{x,y}}\nonumber\\
&\hspace{-5mm}{\textstyle+\frac{1}{32}\tilde{\eta}^{abc}*\Pi^{gIJ}(x)F_{bcIJ}(x)h^{me}(x)\utilde{\eta}_{rm(a}h_{g)s}(x)\varphi^{rs}(x)\frac{\partial}{\partial x^e}\delta^3_{x,y}}-(x\leftrightarrow y),\label{eq54}\\
&\hspace{-25mm}\{\mathcal{H}(x),\varphi^{ab}(y)\}=\left[{\textstyle -\frac{a_1}{4a_2}h_{cf}(x)\tilde{\eta}^{(a|cd}\varphi^{f|b)}(x)\frac{\partial}{\partial x^d}}+\Psi^{ab}\right]\delta^3_{x,y},\label{eq55}\\
&\hspace{-25mm}\{\varphi^{ab}(x),\varphi^{cd}(y)\}=0,\label{eq56}
\end{eqnarray}
 
\noindent where

\begin{equation}\label{eq57}
 \Psi^{ab}\equiv\frac{1}{2}h_{cf}\left(-\tensor{\Pi}{^f_{IJ}}+\frac{\sigma a_1}{2a_2}*\tensor{\Pi}{^f_{IJ}}\right)\tilde{\eta}^{(a|cd}D_d\Pi^{|b)IJ}.
\end{equation}

% We add some comments related to the change \stackrel{(\gamma)}{\Pi} \rightarrow \Pi in section 3

Note that the primary constraint algebra (off-shell) of the CMPR action differs a little from the primary constraint algebra of Sect. 3 (with $\Lambda=0$) when we make there the change $\stackrel{(\gamma)}{\Pi} \rightarrow \Pi$. Firstly, there is an extra term appearing in the term proportional to $H_{arcf}$ of Eq. (\ref{eq30}) regarding the Poisson bracket (\ref{eq52}). Secondly, the term proportional to the derivative of the Dirac's delta in Eq. (\ref{eq55}) does not appear in the Eq. (\ref{eq33}) once we make the above change. On the other hand, the left Poisson brackets have the same form of the corresponding ones of Sect. 3 and we see that (\ref{eq57}) concides with (\ref{eq35}) when we make the mentioned change and identify $4\sigma\gamma^{-1}/(1+\sigma\gamma^{-2})$ with $a_1/a_2$ (but they are not exactly equal because the factor $h^{ab}$ of Sect. 3 differs from that of this section after the mentioned transformation on the canonical momentum is done).

Up to here, the algebra does not close because of (\ref{eq55}) and the theory will have secondary constraints. Since $N\neq 0$, the evolution of the constraint $\varphi^{ab}$ gives rise to the constraint $\Psi^{ab}\approx 0$ once the primary constraints are taken strongly; the other constraints have a trivial evolution.

The requiered Poisson brackets among the new constraint (\ref{eq57}) and the primary constraints are

\begin{eqnarray}
 &\hspace{-25mm}\{\Psi^{ab}(x),\mathcal{G}^{IJ}(y)\}=0,\label{eq58}\\
 &\hspace{-25mm}\{\Psi^{ab}(x),\mathcal{H}_c(y)\}={\textstyle \left[\frac{1}{2}\Psi^{ab}(y)\frac{\partial}{\partial y^c}-\frac{1}{2}\Psi^{ab}(x)\frac{\partial}{\partial x^c}+\delta_c^{(a}\Psi^{b)d}(x)\frac{\partial}{\partial x^d}\right]\delta^3_{x,y}}\nonumber\\
 &\hspace{10mm}{\textstyle +\frac{a_1}{8a_2}\tilde{\eta}^{(a|de}h_{df}(x)\left[\delta_c^m\varphi^{f|b)}(x)-\delta_c^{b)}\varphi^{fm}(x)\right]\frac{\partial^2}{\partial x^e \partial x^m}\delta^3_{x,y}}\nonumber\\
 &\hspace{10mm}{\textstyle +\frac{1}{4}\tilde{\eta}^{(a|de}h_{dc}\left(-D_e\tensor{\Pi}{^{b)}_{IJ}}+\frac{\sigma a_1}{2a_2}*D_e\tensor{\Pi}{^{b)}_{IJ}}\right)\mathcal{G}^{IJ}}\delta^3_{x,y}\nonumber\\
 &\hspace{10mm}{\textstyle+\frac{\sigma}{8h}H_{cndf}\tensor{\Pi}{^n_{IJ}}\left(-\tensor{\Pi}{^f_{KL}}+\frac{\sigma a_1}{2a_2}*\tensor{\Pi}{^f_{KL}}\right)\tilde{\eta}^{(a|de}D_e\Pi^{|b)KL}\mathcal{G}^{IJ}\delta^3_{x,y}}\nonumber\\
 &\hspace{10mm}{\textstyle +\frac{1}{4}\delta_c^{(a}\tilde{\eta}^{b)de}h_{df}(x)\left(-\tensor{\Pi}{^f_{IJ}}(x)+\frac{\sigma a_1}{2a_2}*\tensor{\Pi}{^f_{IJ}}(x)\right)(D_x)_e(G^{IJ}\delta^3_{x,y})},\label{eq59}\\
 &\hspace{-25mm}\{\Psi^{ab}(x),\mathcal{H}(y)\}\approx F^{ab}\delta^3_{x,y},\label{eq60}\\
 &\hspace{-25mm}\{\Psi^{ab}(x),\varphi^{cd}(y)\}=\bar{M}^{(ab)(cd)}\delta^3_{x,y},\label{eq61}
\end{eqnarray}

\noindent where

\begin{equation}
 \hspace{-25mm}\bar{M}^{(ab)(cd)}\equiv{\textstyle\sigma h_{ef}\left(-\tensor{\Pi}{^f_{IJ}}+\frac{\sigma a_1}{2a_2}*\tensor{\Pi}{^f_{IJ}}\right)\left[\left(*\tensor{\Pi}{^{cI}_K}-\frac{a_1}{2a_2}\tensor{\Pi}{^{cI}_K}\right)\tilde{\eta}^{(a|de}\Pi^{|b)KJ}+(c{\scriptstyle\leftrightarrow} d)\right]}
\end{equation}

% here I add compare with (\ref{matrix1})

\noindent is a non-singular matrix (compare with (\ref{matrix1})). To obtain Eq. (\ref{eq60}), we have followed the same steps leading to the Eq. (\ref{eq42}). The evolution of the constraint $\Psi^{ab}$ then allows us to fix the multipliers $\lambda_{ab}$ and the Dirac's method ends. Replacing the constraint $\mathcal{H}$ as in Eq. (\ref{eq44}), we find that the redefined constraint $\bar{\mathcal{H}}$ is first-class, as well as $\mathcal{G}^{IJ}$ and $\mathcal{H}_a$ are. $\varphi^{ab}$ and $\Psi^{ab}$ are second-class because of Eq. (\ref{eq61}), and since we have 18 configuration variables $A_{aIJ}$, the physical local degrees of freedom are $[(2\times18-2(6+3+1)-(6+6)]/2=2$, as expected for a theory describing general relativity.

\section{Concluding remarks}

Using Dirac's method, we have shown that the two action principles (\ref{eq14}) and (\ref{eq45}) describing general relativity as a constrained BF theory and including the Immirzi parameter have two physical degrees of freedom, as expected. The constraint algebra was explicitly calculated in both cases from which we can see the presence of second-class constraints; this fact has already been noted in the Hamiltonian analysis of the Holst action \cite{Barros}, which is the first-order formulation of general relativity with the Immirzi parameter.

% Modifications concerning the comparison in terms of canonical variables

As can be noted above, the constraint algebra of both BF principles of gravity is pretty similar, which is expected since we managed the constraints on the B field in similar ways. However, the constraint algebras are not exactly the same off-shell. Since the best way to compare them is to identify their canonical variables (as it was already done above), we see that by performing the change $\stackrel{(\gamma)}{\Pi}\rightarrow\Pi$ in the constraint algebra of Sect. 3 (with $\Lambda=0$), the resulting algebra looks like the algebra of Sect. 4, but some Poisson brackets get a little modification. For example, we already metioned that there are missing terms when we compare (\ref{eq30}) with (\ref{eq52}) and (\ref{eq33}) with (\ref{eq55}). Also, when we perform this change in (\ref{eq38}), we see that it differs a little from (\ref{eq59}), since the former has an extra term proportional to $H_{cndf}$, but lacks the term proportional to the second derivative of Dirac's delta that the latter has. The origin of these deviations can be tracked to the variable $h^{ab}$ introduced to handle the Eqs. (\ref{eq16}) and (\ref{eq46}), which was defined in terms of $\Pi$ in both cases, but that takes different forms when we express it in terms of the associated canonical momentum (see Eq. (\ref{var3})), and subsequently this difference propagates along all the constraint algebra of Sect. 3. Nevertheless, the constraint algebras of Sects. 3 and 4 coincide on-shell, as expected. In fact, the two constraint algebras can be transformed one into each other by suitably redefining the constraint $\mathcal{H}$: from (\ref{var3})-(\ref{var4}), we see that if we neglect the terms proportional to the constraint $\Phi^{ab}$, then $hh^{ab}\propto \stackrel{(\gamma)}{(hh^{ab})}$. In this way, the factor $h_{ad}$ of equation (\ref{eq21}) becomes $\stackrel{(\gamma)}{h}_{ad}$ and the primary constraints (\ref{eq20})-(\ref{eq23}) become essentially the primary constraints (\ref{cons1})-(\ref{cons4}). This is in full agreement with the fact that both actions principles are equivalent at the Lagrangian level \cite{MontesMer}. Finally, in the case of the BF principle of Sect. 3, the coupling of the cosmological constant only affects the scalar constraint (\ref{eq21}) in the usual way.

%As can be noticed above, the constraint algebra for both BF principles of gravity is pretty similar, though the Immirzi parameter is included there in different ways. More precisely, in Sect. 3 the Immirzi parameter enters in one of the BF terms involved, affecting later the definition of the momentum conjugate to $A_{aIJ}$ and thus being carried by the momentum throughout the Hamiltonian analysis. In fact, there is an ambiguity in choosing the canonical pair to parametrize the phase space of the theory: the pairs $(A,\stackrel{(\gamma)}{\Pi})$ and $(\stackrel{(\gamma)}{A},\Pi)$ are allowed, but they are completely equivalent because the symplectic structure does not change. In section 4, the Immirzi parameter is included through the modification of the constraint on the 2-forms $B^{IJ}$; the result is that only the constraint $\varphi^{ab}$ gets modified (see (\ref{cons2})) regarding the constraint $\Phi^{ab}$ of Sect. 3. In the case of the BF principle of Sect. 3, the coupling of the cosmological constant only affects the scalar constraint (\ref{eq21}) in the usual way.

We want to stress that a canonical analysis of the BF action principle contained in Sect. 3 (without cosmological constant) has also been performed in Ref. \cite{Perez}, but it is not Lorentz-covariant because the time gauge was imposed there, which allows it to recover the real SU(2) formulation of general relativity where second-class constraints do not appear. The results of our Hamiltonian analysis can be useful in two ways. Firstly, we can make contact with the results of \cite{Perez} by solving the second-class constraints, which amounts to parametrize suitably the phase space variables  and perform a time gauge. Secondly, we can maintain the Lorentz-covariant property of the theory and make contact with the formulation of Ref. \cite{Barros}, which initially keeps the second-class constraints and then the Barbero Hamiltonian formulation is obtained by solving them and making the appropriate gauge-fixing.

%In the case of the Lorentz-covariant analysis that we performed in Sect. 3 is expected that the canonical formulation of Ref. \cite{Barros} be recovered once we perform an appropriate parametrization of the phase space variables.

Finally, since both BF action principles for gravity studied in Sects. 3 and 4 possess second-class constraints, handling them by the Dirac bracket method or by solving them explicitly will enable us to make contact with the Lorentz-covariant Hamiltonian formulations for first-order general relativity that include the Immirzi parameter analized in Refs. \cite{Alexandrov1,Alexandrov2} and \cite{Karim,Geiller}. The results of Sects. 3 and 4 can be considered as generalizations of the results of the first part of the work reported in Ref. \cite{Krasn}.

\section{Acknowledgments}

This work was supported in part by CONACYT, México, Grant No. 167477-F.

\end{document}